\documentclass[pra,showpacs,letterpaper,showpacs,superscriptaddress]{revtex4}
\usepackage{appendix}

\begin{document}
\newcommand{\bEp}{\bE^{(+)}}
\newcommand{\bEm}{\bE^{(-)}}
\newcommand{\bEpm}{\bE^{(\pm)}}
\newcommand{\pr}{^{\prime}}
\newcommand{\bd}{{\bf d}}      
\newcommand{\bv}{{\bf v}}
\newcommand{\hbp}{\hat{\bp}}
\newcommand{\hbx}{\hat{\bx}}
\newcommand{\hq}{\hat{q}}
\newcommand{\hp}{\hat{p}}
\newcommand{\ha}{\hat{a}}
\newcommand{\had}{{a}^{\dag}}
\newcommand{\ad}{a^{\dag}}
\newcommand{\hsig}{{\hat{\sigma}}}
\newcommand{\nt}{\tilde{n}}
\newcommand{\itf}{\sl}
\newcommand{\eps}{\epsilon}
\newcommand{\bsig}{\pmb{$\sigma$}}
\newcommand{\beps}{\pmb{$\eps$}}
\newcommand{\bmu}{\pmb{$ u$}}
\newcommand{\balpha}{\pmb{$\alpha$}}
\newcommand{\bbeta}{\pmb{$\beta$}}
\newcommand{\bgamma}{\pmb{$\gamma$}}
\newcommand{\bu}{{\bf u}}
\newcommand{\bpi}{\pmb{$\pi$}}
\newcommand{\bSig}{\pmb{$\Sigma$}}
\newcommand{\be}{\begin{equation}}
\newcommand{\ee}{\end{equation}}
\newcommand{\bea}{\begin{eqnarray}}
\newcommand{\eea}{\end{eqnarray}}
\newcommand{\sss}{_{{\bf k}\lambda}}
\newcommand{\ssss}{_{{\bf k}\lambda,s}}
\newcommand{\dip}{\langle\sigma(t)\rangle}
\newcommand{\dipp}{\langle\sigma^{\dag}(t)\rangle}
\newcommand{\sig}{{\sigma}}
\newcommand{\sigd}{{\sigma}^{\dag}}
\newcommand{\sigz}{{\sigma_z}}
\newcommand{\ra}{\rangle}
\newcommand{\la}{\langle}
\newcommand{\om}{\omega}
\newcommand{\Om}{\Omega}
\newcommand{\pa}{\partial}
\newcommand{\bR}{{\bf R}}
\newcommand{\bx}{{\bf x}}
\newcommand{\br}{{\bf r}}
\newcommand{\bE}{{\bf E}}
\newcommand{\bH}{{\bf H}}
\newcommand{\bB}{{\bf B}}
\newcommand{\bP}{{\bf P}}
\newcommand{\bD}{{\bf D}}
\newcommand{\bA}{{\bf A}}
\newcommand{\bek}{\hat{\bf e}\rmk}
\newcommand{\rmk}{_{{\bf k}\lambda}}
\newcommand{\bsij}{{\bf s}_{ij}}
\newcommand{\bk}{{\bf k}}
\newcommand{\bp}{{\bf p}}
\newcommand{\epso}{{1\over 4\pi\eps_0}}
\newcommand{\BB}{{\mathcal B}}
\newcommand{\AAA}{{\mathcal A}}
\newcommand{\NN}{{\mathcal N}}
\newcommand{\mm}{{\mathcal M}}
\newcommand{\RR}{{\mathcal R}}
\newcommand{\bS}{{\bf S}}
\newcommand{\bL}{{\bf L}}
\newcommand{\bJ}{{\bf J}}
\newcommand{\bI}{{\bf I}}
\newcommand{\bF}{{\bf F}}
\newcommand{\bsub}{\begin{subequations}}
\newcommand{\esub}{\end{subequations}}
\newcommand{\baline}{\begin{eqalignno}}
\newcommand{\ealine}{\end{eqalignno}}
\newcommand{\isat}{{I_{\rm sat}}}
\newcommand{\Is}{I^{\rm sat}}
\newcommand{\Ip}{I^{(+)}}
\newcommand{\Imm}{I^{(-)}}
\newcommand{\Inu}{I_{\nu}}
\newcommand{\bInu}{\overline{I}_{\nu}}
\newcommand{\bN}{\overline{N}}
\newcommand{\qnu}{q_{\nu}}
\newcommand{\oqn}{\overline{q}_{\nu}}
\newcommand{\qsat}{q^{\rm sat}}
\newcommand{\Iout}{I_{\nu}^{\rm out}}
\newcommand{\topt}{t_{\rm opt}}
\newcommand{\crr}{{\mathcal{R}}}
\newcommand{\cE}{{\mathcal{E}}}
\newcommand{\cH}{{\mathcal{H}}}
\newcommand{\epsoo}{\epsilon}
\newcommand{\ombar}{\overline{\om}}
\newcommand{\cEp}{{\mathcal{E}}^{(+)}}
\newcommand{\cEm}{{\mathcal{E}}^{(-)}}
\newcommand{\Ep}{E^{(+)}}
\newcommand{\Em}{E^{(-)}}
\newcommand{\ppp}{p^{(+)}}
\newcommand{\ppm}{p^{(-)}}
\newcommand{\Ed}{E^{\dag}}
\newcommand{\pd}{p^{\dag}}
\newcommand{\TT}{{\mathcal{T}}}
\title{Free Energy of Coupled Oscillators: Lamb Shifts and van der Waals Interactions}
\author{Peter W. Milonni}
\affiliation{Department of Physics and Astronomy, University of Rochester, Rochester, New York 14627 USA}
\begin{abstract}
The Helmholtz free energy of oscillators in thermal equilibrium with electromagnetic radiation is obtained from the Pauli--Hellmann--Feynman theorem and applied to some aspects of Lamb shifts and van der Waals interactions.
\end{abstract}
\maketitle
\section{Introduction}
Lamb shifts and van der Waals interactions may be attributed to the coupling of atoms to the zero-point electromagnetic field. These effects are modified at finite temperatures and depend on the mode structure of the field. Analyses of  these effects have involved different formalisms and physical interpretations, all based in one way or another on quantum fluctuations of electromagnetic fields and their sources, and many invoking in particular the zero-point energy of the field or its finite-temperature generalization. Here we take an approach based on the Pauli--Hellmann--Feynman (PHF) theorem. We begin in 
Section \ref{sec:energy} with brief, heuristic derivations of the (nonrelativistic) Lamb shift and the van der Waals interaction based on changes in zero-point field energy. In Section \ref{sec:phf} we use the PHF theorem to derive an exact expression for the Helmholtz free energy of a system coupled to a heat bath, including many-body interactions. This is then applied in the following sections to some aspects of Lamb shifts and van der Waals interactions, and in particular to the form of the van der Waals interaction when there is strong coupling to a single field mode. The physical interpretation of these results is briefly discussed in Section \ref{sec:remarks}.

\section{\label{sec:energy} Scatterings: Lamb Shift and van der Waals Interaction at Zero Temperature}
Sixty years ago, in a talk at the Relativity Conference in Warsaw, Richard Feynman \cite{feyn63} returned to an interpretation of the hydrogen Lamb shift he had suggested earlier \cite{feyn61,power}. The argument, briefly, is as follows. In a box of volume $V$ containing $N$ identical atoms per unit volume, the zero-point energy of a field mode of frequency $\om$ is $\frac{1}{2}\hbar\om/n(\om)$, where $n(\om)$ is the refractive index. The change in the total zero-point energy due to the presence of the atoms is therefore 
\be\Delta E=2V\int \frac{d^3k}{(2\pi)^3}\frac{1}{2}\hbar\om\big(\frac{1}{n}-1\big)\cong -\frac{\hbar}{\pi c^3}\int_0^{\infty}d\om\om^3\alpha_0(\om)
\label{eq1}
\ee
in the case of a single atom ($NV=1$) with polarizability $\alpha_0(\om)$, $n(\om)\cong 1 +2\pi N\alpha_0(\om)$. If we use the Kramers--Heisenberg formula for $\alpha_0(\om)$, subtract out the free-electron energy given by $\Delta E$ with $\alpha_0(\om)=-e^2/m\om^2$, and introduce a high-frequency cutoff $mc^2/\hbar$, we obtain, without any need for mass renormalization, exactly the ``Bethe log" expression  for the (nonrelativistic) Lamb shift \cite{bethe,jordan}. 
This is discussed in a bit more detail in Section \ref{sec:lamb}.

The formula (\ref{eq1}) can be expressed in terms of the forward scattering amplitude $f(\om)=\alpha_0(\om)\om^2/c^2$:
\be
\Delta E=-2\pi\hbar c^2\int\frac{d^3k}{(2\pi)^3}\frac{f(\om)}{\om},
\label{eq2}
\ee
which is essentially Feynman's formula \cite{feyn61,remark1}. It is equivalent to Bethe's, but it involves a scattering amplitude for a {\sl real} scattering process, whereas Bethe's formula involves the single closed-loop diagram for emission and absorption of  {\sl virtual} photons. Feynman remarked that the formula 
(\ref{eq2}) is simple but ``very peculiar. The reason it's peculiar is that the forward scatterings are real processes. At last I had discovered a formula I had always wanted, which is a formula for energy differences (which are defined in terms of virtual fields) in terms of actual measurable quantities ..." \cite{feyn63}.

A more direct calculation for an atom at position $\br_A$ leads to
\be
\Delta E(\br_A)=-\frac{1}{2}\sum\rmk\big|\big(\frac{2\pi\hbar\om}{V}\big)^{1/2}\bek e^{i\bk\cdot\br_A}\big|^2\alpha_0(\om)=-\frac{\hbar}{8\pi^2}\sum_{\lambda=1}^{2}\int d^3k\om\big|\bek e^{i\bk\cdot\br_A}\big|^2\alpha_0(\om),
\label{eq3}
\ee
which of course is equivalent to Eq. (\ref{eq2}). Here $\bek$ is a linear polarization unit vector ($\bk\cdot\bek=0, \lambda=1,2$). Now suppose there is an identical atom B at a position $\br_B$, both atoms in their ground states. The effect on atom A is to replace $\bek e^{i\bk\cdot\br_A}$  by
\be
\bek e^{i\bk\cdot\br_A}+\alpha_0(\om)e^{i\bk\cdot\br_B}k^3e^{ikr}\Big\{\big[\bek-(\bek\cdot\hat{\br})\hat{\br}\big]\frac{1}{kr}+\big[\bek-3(\bek\cdot\hat{\br})\hat{\br}\big]\big[\frac{i}{k^2r^2}-\frac{1}{k^3r^3}\big]\Big\},
\label{eq4}
\ee
where $\br=\br_A-\br_B$, $\hat{\bf r}=\br/r$, and $\bk=k\hat{\bk}$. The second term may be thought of as the field at A from the dipole moment induced in B by the vacuum field incident on B, i.e., it may be attributed to scattering of the vacuum field by atom B. When we use this expression in place of  $\bek e^{i\bk\cdot\br_A}$ in Eq. (\ref{eq3}), and retain only terms up to second order in $\alpha_0(\om)$, we obtain, in addition to the $r$-independent Lamb shift of atom A, an $r$-dependent energy
\be
\Delta E(r)=-\frac{\hbar c}{\pi r^2}\int_0^{\infty}duu^4\alpha_0^2(icu)\Big(1+\frac{2}{ur}+\frac{5}{u^2r^2}+\frac{6}{u^3r^3}+\frac{3}{u^4r^4}\Big)e^{-2ur},
\label{eq6}
\ee
a well-known expression for the van der Waals interaction of two molecules in a vacuum, neither of which has a permanent dipole moment. (We have used the analyticity of $\alpha_0(\om)$ in the first quadrant of the complex frequency plane to analytically continue the integral along the positive real axis to an integral along the positive imaginary axis.) In the limit of very large separations this gives the Casimir--Polder result, $\Delta E(r)=-23\hbar c\alpha_0^2(0)/4\pi r^7$, for the retarded van der Waals interaction, whereas at small separations it gives the London result in which $\Delta E(r)\propto  1/r^6$. Like the Lamb shift, the van der Waals interaction can be expressed in terms of a {\sl real} scattering process and a forward scattering amplitude. The zero-point field is Rayleigh-scattered by each atom according to the expression (\ref{eq4}), and the scattered field modifies the zero-point field at the other atom from its free-space form, resulting, in effect, in an $r$-dependent Lamb shift. This is the van der Waals interaction energy. The extension to many-atom systems, multiple scattering, and finite temperatures is perhaps most easily done with a simple extension of the Pauli--Hellmann--Feynman theorem (Section \ref{sec:phf}).

\section{\label{sec:phf} Free Energy of Atoms in Thermal Equilibrium with Radiation}
\subsection{Pauli--Hellmann--Feynman theorem for free energy}
Consider a Hamiltonian of the general  form $H=H_0+ \lambda H_1$, where $H_0$ is the unperturbed Hamiltonian and the interaction Hamiltonian is parametrized by a coupling constant $\lambda$.  The eigenvalues $E(\lambda$) and eigenvectors $|\psi(\lambda)\ra$ of $H$ will of course depend on $\lambda$. According to the Pauli--Hellmann-Feynman theorem \cite{hellfeyn,musher}, 
\be
\frac{dE}{d\lambda}=\la\psi(\lambda)|\frac{dH}{d\lambda}|\psi(\lambda)\ra.
\label{eq7}
\ee
In its integral form the PHF theorem gives the change $E(1)-E(0)$ in the energy of the system in the form of the coupling-constant integration algorithm:
\be
E(1)-E(0)=\int_0^1 \frac{d\lambda}{\lambda} \la \psi(\lambda)|\lambda H_1|\psi(\lambda)\ra,
\label{eq8}
\ee
the difference between the energy with ($\lambda=1$) and without ($\lambda=0$) the interaction $H_1$ \cite{sawada}.

In the case of thermal equilibrium there is an expression analogous to Eq. (\ref{eq7}), now involving the average $\la ... \ra$ over the canonical ensemble, for the Helmholtz free energy $F(\lambda,T)$  \cite{paul,pons}:
\be
\frac{dF}{d\lambda}=\Big\la\frac{dH}{d\lambda}\Big\ra,
\label{eq9}
\ee
which follows simply from the definition
\be
F(\lambda,T)=-k_BT\ln{\rm Tr}\Big[e^{-H(\lambda)/k_BT}\Big].
\label{eq10}
\ee
Integration of Eq. (\ref{eq9}) gives the change in the free energy in a form similar to the zero-temperature expression (\ref{eq8}):
\be
\Delta F=F(1,T)-F(0,T)=\int_0^1 \frac{d\lambda}{\lambda}\la \lambda H_1\ra.
\label{eq11}
\ee
\subsection{\label{sec:coupling}Coupling of induced dipoles and thermal radiation}
We consider now a collection of $\NN$ atoms coupled to a heat bath, specifically an electromagnetic field in thermal equilibrium at temperature $T$. The atoms are assumed to remain in their ground states with high probability. We assume there are no permanent dipole moments, only electric dipole moments induced by the field. The interaction Hamiltonian in the electric dipole approximation is
\be
H_{\rm int}=-\frac{1}{2}\sum_{n=1}^{\NN}\sum_{i=1}^3\big[p_i(\br_n,t)E_i(\br_n,t) +E_i(\br_n,t)p_i(\br_n,t)\big],
\label{eq13}
\ee
where $E_i(\br_n,t)$ is the $i$th component of the electric field operator for the thermal field at the position $\br_n$ of the atom with dipole moment $\bp(\br_n,t)$.  Effects of fields from the atoms themselves are subsumed in the polarizability, as done below. The coupling constant for the application of the PHF theorem is the electron charge $e$.  We write $E_i(\br_n,t)$ in terms of positive- and negative-frequency components as
\be
E_i(\br_n,t)=\int_0^{\infty}d\om\big[\Ep_i(\br_n,\om)e^{-i\om t}+\Em_i(\br_n,\om)e^{i\om t}\big],
\label{eq14}
\ee
and likewise 
\be
p_i(\br_n,t)=\int_0^{\infty}d\om\big[\ppp_i(\br_n,\om)e^{-i\om t}+\ppm_i(\br_n,\om)e^{i\om t}\big],
\label{eq16}
\ee
with \cite{remark2}
\be
\ppp_i(\br_n,\om)=\alpha_0(\om+i0^+)E_i(\br_n,\om)
\label{eq17}
\ee
in the case of a single atom. The polarizability $\alpha_0(\om+i0^+)$  is given by the Kramers--Heisenberg formula:
\be
\alpha_0(\om+i0^+)=\frac{2}{3\hbar}\sum_s\frac{\om_{sg}|\bd_{sg}|^2}{\om_{sg}^2-(\om+i0^+)^2},
\label{eq18}
\ee
where $\om_{sg}$ ($>0$) is the frequency for the transition between the ground state $g$ and the excited state $s$ and $\bd_{sg}$ is the corresponding electric dipole matrix element. For $\NN$ atoms the dipole moment induced in any one atom is 
\be
\ppp_i(\br_n,\om)=\alpha_0(\om+i0^+)\Ep_i(\br_n,\om)+\alpha_0(\om+i0^+)\sum_{m=1}^{\NN}\sum_{j=1}^3G_{ij}(\br_n,\br_m,\om)\ppp_j(\br_m,\om),
\label{ad1}
\ee
where the dyadic Green function $G(\br_n,\br_m,\om)$ is defined in the Appendix. In matrix form, 
\be
\ppp(\om)=\alpha_0(\om+i0^+)\Ep(\om)+\alpha_0(\om+i0^+)G(\om)\ppp(\om),
\label{ad2}
\ee
or
\be
\ppp(\om)=\frac{\alpha_0(\om+i0^+)\Ep(\om)}{1-\alpha_0(\om+i0^+)G(\om)}\equiv\alpha(\om)\Ep(\om),
\label{ad3}
\ee
where $\alpha(\om)$ and $G(\om)$ are $3\NN\times 3\NN$ matrices and $\ppp(\om)$ is a $3\NN$-dimensional vector.

For thermal radiation the different frequency components of $\bE(\br,t)$ are uncorrelated. As reviewed in the Appendix,
\bea
\big\la \Ep_i(\br_n,\om)\Em_j(\br_m,\om\pr)\big\ra&=&\frac{\hbar}{\pi}[q(\om)+1][G^I_{ij}(\br_n,\br_m,\om)]\delta(\om-\om\pr),\nonumber\\
\big\la \Em_i(\br_n,\om)\Ep_j(\br_m,\om\pr)\big\ra&=&\frac{\hbar}{\pi}q(\om)[G^I_{ij}(\br_n,\br_m,\om)]\delta(\om-\om\pr),
\label{eq19}
\eea
where $G^I_{ij}(\br_n,\br_m,\om) $ is the imaginary part of $G_{ij}(\br_n,\br_m,\om)$ ($=G_{ji}(\br_m,\br_n,\om)$) and $q(\om)=(e^{\hbar\om/k_BT}-1)^{-1}$. Thus, 
\bea
\la H_{\rm int}\ra&=&-\frac{\hbar}{\pi}\sum_{n,m=1}^{\NN}\sum_{i,j=1}^3\int_0^{\infty}d\om\alpha_{ij}(\br_n,\br_m,\om)[2q(\om)+1]G_{ji}^I(\br_m,\br_n,\om)
=-\frac{\hbar}{\pi}{\rm Im}{\rm Tr}\int_0^{\infty}d\om\alpha(\om)[2q(\om)+1]G(\om)\nonumber\\
&=&-\frac{\hbar}{\pi}{\rm Im}{\rm Tr}\int_0^{\infty}d\om\frac{\alpha_0(\om+i0^+)G(\om)}{1-\alpha_0(\om+i0^+)G(\om)}\coth\big(\frac{\hbar\om}{2k_BT}\big).
\label{eq20}
\eea

\subsection{\label{sec:free}Free energy}
As noted above, the coupling constant for the application of the PHF theorem may be taken to be the electron charge $e$. Since $\alpha_0(\om+i0^+)$ is proportional to $e^2$, it follows from Eq. (\ref{eq20}) that
\bea
\Delta F&=&-\frac{\hbar}{\pi}{\rm Im}{\rm Tr}\int_0^{\infty}d\om\coth\big(\frac{\hbar\om}{2k_BT}\big)\int_0^1\frac{d\lambda}{\lambda}\frac{\lambda^2\alpha_0(\om+i0^+)G(\om)}
{1-\lambda^2\alpha_0(\om+i0^+)G(\om)}\nonumber\\
&=&\frac{\hbar}{2\pi}{\rm Im}{\rm Tr}\int_0^{\infty}d\om\log\big[1-\alpha_0(\om+i0^+)G(\om)\big]\coth\big(\frac{\hbar\om}{2k_BT}\big).
\label{eqfree}
\eea
Using the identity ${\rm Tr}\log[1-X]=\log\det(1-X)$, we can write this as 
\be 
\Delta F(T)=-\frac{\hbar}{2\pi}{\rm Im}\int_0^{\infty}d\om\log\det\Big[\frac{\alpha(\om)}{\alpha_0(\om+i0^+)}\Big]\coth\big(\frac{\hbar\om}{2k_BT}\big),
\label{eqfr2}
\ee
which has the form of  the multi-particle generalization of the ``remarkable theorem" of Ford, Lewis, and O'Connell \cite{ford1} when we identify $\alpha(\om)=\alpha_0(\om+i0^+)/[1-\alpha_0(\om+i0^+)G(\om)]$ as their ``generalized susceptibility." This formula gives the Helmholtz free energy of the {\sl interacting} system of oscillators, in this case atoms and the electromagnetic field, in terms of the polarizability of the atoms alone. A different derivation is given in the original paper of Ford et al. \cite{ford1}. Another derivation, based essentially on the PHF theorem but not in the form of the coupling-constant integration algorithm used here, is given in Reference \cite{paul}. 

\section{\label{sec:lamb} Lamb shifts}
Retaining only the term linear in $\alpha_0(\om+i0^+)$, Eq. (\ref{eqfree}) gives, for a single atom at any point $\br$ in free space,
\be
\Delta F(T)=-\frac{\hbar}{2\pi}{\rm Im}\sum_{i=1}^3\int_0^{\infty}d\om\alpha_0(\om+i0^+)G_{ii}(\br,\br,\om)\coth\big(\frac{\hbar\om}{2k_BT}\big)=
-\frac{\hbar}{\pi c^3}\int_0^{\infty}d\om\om^3\alpha_0(\om+i0^+)\coth\big(\frac{\hbar\om}{2k_BT}\big),
\label{eq100}
\ee
since, from Eq. (\ref{eq27}), Im$\big[\lim_{\br\rightarrow\br\pr}G(\br,\br\pr,\om)\big]=2\om^2k/c^2=2\om^3/c^3$. For $T=0$ this reproduces Eq. (\ref{eq1}). Subtracting the free-electron ($\om_{sg}\rightarrow 0$) contribution, and introducing a high-frequency cutoff $\Om$, we replace Eq. (\ref{eq100}) by the ``observable" shift $\Delta F(0)_{\rm obs}$, i.e., the difference in the shift between bound and unbound electrons:
\bea
\Delta F(0)_{\rm obs}&=&-\frac{2}{3\pi c^3}{\rm P}\int_0^{\Om}d\om\om^3\sum_s\om_{sg}|\bd_{sg}|^2\Big(\frac{1}{\om^2_{sg}-\om^2}-\frac{1}{-\om^2}}\Big)
=-\frac{2}{3\pi c^3}\sum_s\om^2_{sg}{\bd_{sg}|^2{\rm P}\int_0^{\Om}\frac{\om d\om}{\om^2_{sg}-\om^2}\nonumber\\
&=&-\frac{2}{3\pi c^3}\sum_s\om^2_{sg}|\bd_{sg}|^2\int_0^{\Om}\frac{d\om}{\om+\om_{sg}}
\label{bethe}
\eea
for $\Om\gg |\om_{sg}|$ for all transition frequencies $\om_{sg}$. (P stands for ``principal part.") This of course is the ``Bethe log" when we take the high-frequency cutoff $\Om$ to be $mc^2/\hbar$. 

For an atom in a homogeneous dielectric medium, Im$\big[\lim_{\br\rightarrow\br\pr}G(\br,\br\pr,\om)\big]=2n(\om)\om^3/c^3$ and
\be
\Delta F(0)_{\rm diel}=-\frac{2}{3\pi c^3}\sum_s\om^2_{sg}|\bd_{sg}|^2\int_0^{\Om}\frac{n(\om)d\om}{\om_{sg}+\om}.
\ee
The difference between the Lamb shift of an atom in the dielectric and the atom in vacuum is
\be
\Delta F(0)_{\rm diel}-\Delta F(0)_{\rm vac}=-\frac{2}{3\pi c^3}\sum_s\om^2_{sg}|\bd_{sg}|^2\int_0^{\Om}\frac{[n(\om)-1]d\om}{\om_{sg}+\om}.
\ee
Since $n(\om)-1$ can be expected to vary as $1/\om^2$ as $\om\rightarrow\infty$, we can take $\Om\rightarrow\infty$. In any event, it appears that this modified Lamb shift  would be very difficult to observe because of competing effects and shifts resulting from the interaction of the guest atom with the host atoms of the medium.

Equation (\ref{eq100})  implies a $T$-dependent correction to the Lamb shift:
\be
\Delta F_i(T)-\Delta F_i(0)=-\frac{4}{3\pi c^3}\sum_j|\bd_{ij}|^2{\rm P}\int_0^{\infty}\frac{d\om\om^3}{e^{\hbar\om/k_BT}-1}\frac{\om_{ji}}
{\om_{ji}^2-(\om+i0^+)^2}
\ee
for an atom in state $i$. For transition frequencies and temperatures such that $\hbar|\om_{ji}|\ll k_BT$ \cite{wing},
\be
\Delta F_i(T)-\Delta F_i(0)\cong \frac{4}{3\pi c^3}\sum_j|\bd_{ij}|^2\om_{ji}\int_0^{\infty}\frac{d\om\om}{e^{\hbar\om/k_BT}-1}=\frac{\pi e^2}{3m\hbar c^3}(k_BT)^2,
\label{hall}
\ee
where we have used the Thomas--Reiche--Kuhn sum rule. This is just the average kinetic energy obtained from the equation of motion $m\ddot{\bx}=e\bE$ for an electron in a blackbody field at temperature $T$. Temperature-dependent corrections to the Lamb shift of Rydberg atoms have been measured and found to be consistent with a $T^2$ scaling \cite{hall}.

\section{\label{sec:vdw} Van der Waals Interactions}
The polarizability $\alpha(\om)$ is required from causality considerations to be analytic in the upper half of the complex frequency plane. From the definition of $G(\om)$ it is clear that $\alpha_0(\om+i0^+)G(\om)$ is analytic in the upper half of the complex frequency plane. Assuming for now that $\log[1-\alpha_0(\om+i0^+)G(\om)]$ is likewise analytic, we can analytically continue the integral in Eq. (\ref{eqfree}) and express the (free) energy for $T=0$ as
\be
\Delta F=\frac{\hbar}{2\pi}{\rm Tr}\int_0^{\infty}d\xi\log[1-\alpha_0(i\xi)G(i\xi)].
\label{g1}
\ee
Considering only the contribution that goes as $\alpha_0^2(i\xi)$, and ignoring the self-energy terms with $\br_n=\br_m$, we obtain
\be
\Delta F_2=\sum_{n=1}^{\NN}\sum_{m=1}^{\NN}(1-\delta_{mn})\Big\{-\frac{\hbar}{4\pi}{\rm Tr}\int_0^{\infty}d\xi\alpha_0^2(i\xi)G_{ij}(\br_n,\br_m,i\xi)G_{ji}(\br_m,\br_n,i\xi)\Big\},
\label{g2}
\ee 
which is found to be just the sum of pairwise van der Waals interaction energies given by Eq. (\ref{eq6}). In particular, for small separations the (nonretarded) van der Waals interaction between two ground-state atoms with polarizabilities $\alpha_1(\om)$ and $\alpha_2(\om)$  has the well-known form originally obtained by London:
\bea
\Delta E&=&-\frac{3\hbar}{\pi r^6}\int_0^{\infty}d\xi\alpha_1(i\xi)\alpha_2(i\xi)=-\frac{3\hbar}{\pi r^6}\big(\frac{2}{3\hbar}\big)^2\sum_m\sum_n|\bd_{1m}|^2|\bd_{2n}|^2\om_{1m}\om_{2n}\int_0^{\infty}\frac{d\xi}{(\om_{1m}^2+\xi^2)(\om_{2n}^2+\xi^2)}\nonumber\\
&=&-\frac{2}{3\hbar r^6}\sum_m\sum_n\frac{|\bd_{1m}|^2|\bd_{2n}|^2}{\om_{1m}+\om_{2n}},
\eea
where $\om_{\mu n}$ ($\mu=1,2$) are the transition frequencies between the ground state and the state $n$ and $\bd_{\mu n}$ are the corresponding transition moments. More generally Eq. (\ref{g1}) accounts for many-body interactions and retardation.

It may be worth noting that, since the magnitude of the static polarizability $\alpha_{st}$ is roughly on the order of an atomic radius, we require that $\alpha_{1st}\alpha_{2st}/r^6=\alpha_{1st}\alpha_{2st}G_{ij}(\br_1,\br_2,\om)G_{ji}(\br_2,\br_1,\om)<1$ for small $r=|\br_1-\br_2|$; otherwise overlap of the atomic wavefunctions must be considered, which we have not done. This condition can also be understood from the requirement that the Hamiltonian must be bounded from below \cite{ford2}.

Renne \cite{renne} obtained a formula similar to Eq. ({\ref{g1}) based on the zero-point energy of coupled harmonic oscillators, each having a frequency $\om_0$. Consider Eq. (\ref{ad1}) without the first term on the right-hand side and without allowing for the coupling of each oscillator to its own field:
\be
\ppp_i(\br_n,\om)=\alpha_0(\om+i0^+)\sum_{m\neq n}^{\NN}G_{ij}(\br_n,\br_m,\om)\ppp_j(\br_m,\om),
\ee
or, in matrix form,
\be
[1+\alpha_0(\om+i0^+)\TT(\om)]\ppp(\om)=0,
\label{g3}
\ee
where $\TT_{ij}(\br_n,\br_m,\om)=-(1-\delta_{mn})G_{ij}(\br_n,\br_m,\om)$. The condition for a non-trivial solution of this set of $\NN$ equations is that the ``normal-mode" frequencies $\om$ must satisfy
\be
f(\om)=\det[1+\alpha_0(\om+i0^+)\TT(\om)]=0.
\label{g4}
\ee
Solutions of this equation in which all $\NN$ values of $\om$ are real can be obtained in the nonretarded regime. In this case Renne has used the argument theorem to obtain the sum of the zeros $\om_s$ of $f(\om)$, and he identifies $\sum_s\frac{1}{2}\hbar\om_s$ as the zero-point energy of the system of oscillators coupled to each other by their electrostatic dipole interactions. The difference $\Delta E$ between this zero-point energy and the zero-point energy $\frac{3}{2}\NN \hbar\om_0$ of the uncoupled oscillators is shown to be 
\be
\Delta E=\frac{\hbar}{2\pi}\int_0^{\infty}d\xi\log\det[1+\alpha_0(i\xi)\TT]=\frac{\hbar}{2\pi}{\rm Tr}\int_0^{\infty}d\xi
\log[1+\alpha_0(i\xi)\TT],
\label{g5}
\ee
which is very similar to Eq. (\ref{g1}) except that self-interactions are excluded. Renne proceeds to generalize this expression to allow for retardation, and his result is equivalent, except for Lamb shifts, to Eq. (\ref{eqfree}) with $T=0$. (The temperature dependence of van der Waals interactions has been studied by several authors \cite{ford2}.)

\section{\label{sec:strong}Strong Coupling of Molecules to a Single Cavity Mode}
There has recently been much interest in modifications of molecular interactions when there is strong coupling of the molecules to a single cavity mode. Haugland et al., for instance, have shown in nonperturbative numerical studies that the distance dependence of van der Waals interactions is significantly affected by such coupling \cite{haugland,philbin}. They also present an illustrative perturbation-theoretic approach based on a Hamiltonian that includes the short-distance intermolecular dipole-dipole interaction
\be
V_{AB}=-\frac{1}{r^3}\big[\bd_A\cdot\bd_B-3(\bd_A\cdot\hat{\br})(\bd_B\cdot\hat{\br})\big]
\label{haug1}
\ee
between molecules A and B which are assumed to have no permanent dipole moments. The alteration of the van der Waals interaction occurs as a result of the additional coupling of the molecules to the vacuum single-mode field. This follows from the PHF theorem, as we now show with a model of $\NN$ two-state atoms interacting with a vacuum single-mode cavity field of frequency $\om$ and polarization $\hat{\bf e}$, and with each other via
\be
V=-\sum_{n=1}^{\NN}\sum_{m=1}^{\NN}\frac{1}{r_{nm}^3}\big[\bd_n\cdot\bd_m-3(\bd_n\cdot\hat{\br}_{nm})(\bd_m\cdot\hat{\br}_{nm})\big]\big[(\sig_n+\sigd_n)(\sig_m+\sigd_m)\big]=-\sum_{n=1}^{\NN}\sum_{m=1}^{\NN}V_{nm}\big[(\sig_n+\sigd_n)(\sig_m+\sigd_m)\big],
\label{haug2}
\ee
where $r_{nm}=|\br_n-\br_m|$ and $\sig_{n}$ and $\sigd_{n}$ are respectively the two-state lowering and raising operators. The transition frequencies and dipole matrix elements of the atoms are denoted by $\om_n$ and $\bd_{n}$. The Hamiltonian for the interaction of the atoms with the cavity field in the electric dipole approximation has the form
\be
H_c=-\sum_{n=1}^{\NN}C_n(a+\ad)(\sig_n+\sigd_n),
\label{haug3}
\ee
where
\be
C_{n}=A_{n}(\bd_{n}\cdot\hat{\bf e})\sqrt{\hbar\om}
\label{haug4}
\ee
and $A_{n}$ is a mode function that depends on the position $\br_n$ of atom $n$ in the cavity.
The complete Hamiltonian is 
 \be
 H=\sum_{n=1}^{\NN}\hbar\om_n\sigd_n\sig_n+\hbar\om\ad a+H_c+V.
 \label{h1}
 \ee
 
 We proceed as in Section \ref{sec:coupling}. The solution of the Heisenberg equation of motion for $\sig_n(t)$, omitting the freely evolving part that plays no role in what follows, is
 \be
 \sig_n(t)=\frac{i}{\hbar}C_n\int_{-\infty}^tdt\pr\big[a(t\pr)+\ad(t\pr)\big]e^{i\om_n(t\pr-t)}
 +\frac{i}{\hbar}\sum_{m}V_{nm}\int_{-\infty}^tdt\pr\big[\sig_m(t\pr)+\sigd_m(t\pr)\big]e^{i\om_n(t\pr-t)}.
 \label{h2}
 \ee
 Since ground-state atoms can be treated effectively as harmonic oscillators for our purposes, we have assumed the commutation relation $[\sig_{\mu}(t),\sigd_{\nu}(t)]=\delta_{\mu\nu}$. Now to lowest order in the coupling constants,
 \be
 a(t\pr)\cong a(t)e^{-i\om(t\pr-t)} \ \ \ \ {\rm and} \ \ \ \ \sig_m(t\pr)\cong \sig_m(t)e^{-i\om_m(t\pr-t)}.
 \label{h3}
 \ee
 It then follows from Eq. (\ref{h2}) and some simple algebra that
 \be
 \sig_{xn}(t)\cong \mathcal{E}_n(t) +\sum_{m=1}^{\NN}\mathcal{V}_{nm}\sig_{xm}(t),
 \label{h4}
 \ee
 where we have defined
 \be
 \mathcal{E}_n(t)=\frac{2C_n}{\hbar}\frac{\om_n}{\om_n^2-\om^2}\big[a(t)+\ad(t)\big],
 \label{h5}
 \ee
 \be
 \mathcal{V}_{nm}=\frac{2}{\hbar}V_{nm}\frac{\om_n}{\om_n^2-\om_m^2}
 \label{h6}
 \ee
 and $\sig_{xn}=\sig_n+\sigd_n$.
 
 From the Heisenberg equation of motion for $a(t)$,
 \bea
 a(t)&=&a_0(t)+\frac{iC_1}{\hbar}\int_{-\infty}^tdt\pr\big[\sig_1(t\pr)+\sigd_1(t\pr)\big]e^{i\om(t\pr-t)}+\frac{iC_2}{\hbar}\int_{-\infty}^tdt\pr\big[\sig_2(t\pr)+\sigd_2(t\pr)\big]e^{i\om(t\pr-t)}\nonumber\\
&\cong& a_0(t)+\frac{C_1}{\hbar}\Big[\frac{\sig_1(t)}{\om-\om_1}+\frac{\sigd_1(t)}{\om+\om_1}\Big]+\frac{C_2}{\hbar}\Big[\frac{\sig_2(t)}
{\om-\om_2}+\frac{\sigd_2(t)}{\om+\om_2}\Big]
 \label{haug6}
 \eea
 in the approximation $\sig_{\mu}(t\pr)\cong\sig_{\mu}(t)e^{-i\om_{\mu}(t\pr-t)}$, with $a_0(t)$ the freely evolving annihilation operator for the single-mode cavity field. Likewise
 \be
 \sig_1(t)\cong\sig_{10}(t)+\frac{C_1}{\hbar}\big[\frac{a_0(t)}{\om_1-\om}+\frac{\ad_0(t)}{\om_1+\om}\Big]-\frac{V_{12}}{\hbar}\Big[\frac{\sig_2(t)}{\om_1-\om_2}+\frac{\sigd_2(t)}{\om_1+\om_2}\Big],
 \label{hh2}
 \ee
 and
 \be
 \sig_2(t)\cong\sig_{20}(t)+\frac{C_2}{\hbar}\Big[\frac{a_0(t)}{\om_2-\om}+\frac{\ad_0(t)}{\om_2+\om}\Big]-\frac{V_{12}}{\hbar}\Big[\frac{\sig_1(t)}{\om_2-\om_1}+\frac{\sigd_1(t)}{\om_1+\om_2}\Big],
 \label{hh3}
 \ee
with $\sig_{\mu 0}(t)$ the freely evolving lowering operator for atom $\mu$. 

For the application of the PHF theorem we require the expectation value of $H_c$  for the state $|\psi\ra$ in which the atoms are in their ground states and the cavity field is in its vacuum state. Considering only atom 1, for instance,
 \be
 \la H_c^{(1)}\ra=-C_1\la\sigd_1a+\ad\sig_1\ra-C_1\la a\sig_1+\sigd_1\ad\ra,
 \label{haug5}
 \ee
 where we have used the fact that the atom and field operators commute, as assumed in writing the Heisenberg equations. Consider first the first term on the right-hand side of Eq. (\ref{haug5}). Since $a_0(t)|\psi\ra=0$, the only nonvanishing part of this term would have to come from the last two terms on the right-hand side of the expression  (\ref{haug6}). But these do not contribute to $-C_1\la\sigd_1a+\ad\sig_1\ra$ any terms involving $\la\sig_{\mu}(t)\sigd_{\mu}(t)\ra =1$ for ground-state atoms, only terms such as $\la\sig_1(t)\sig_1(t)\ra=\la\sig_1(t)\sigd_2(t)\ra=0$. Thus the first term on the right-hand side of Eq. (\ref{haug5}) vanishes within the approximations we have made, and so
 \be
 \la H_c^{(1)}\ra=-C_1\la a\sig_1(t)+\sigd_1(t)\ad(t)\ra\cong -2C_1\la a(t)\sig_1(t)\ra.
\label{hh4}
\ee

We will make the simplifying assumption in this illustrative model that $\om\gg\om_1,\om_2$, as would be the case, for instance, for infrared transitions in an optical cavity. Then
\be
\sig_2(t)\cong\sig_{20}(t)-\frac{C_2}{\hbar\om}[a_0(t)-\ad_0(t)]-\frac{V_{12}}{\hbar}\Big[\frac{\sig_1(t)}{\om_2-\om_1}+\frac{\sigd_1(t)}{\om_2+\om_1}\Big],
\label{hh5}
\ee
and, from the expression (\ref{h2}),
\be
\sig_1(t)\cong\sig_{10}(t)+\frac{C_1}{\hbar}\Big[\frac{a_0(t)}{\om_1-\om}+\frac{\ad_0(t)}{\om_1+\om}\Big]
-\frac{C_2V_{12}}{\hbar\om}\frac{2\om_2}{\om_1^2-\om_2^2}\big[a_0(t)-\ad_0(t)\big]
\label{hh6}
\ee
to first order in $V_{12}$. Thus,
\be
\la a(t)\sig_1(t)\cong\la a_0(t)\sig_1(t)\ra=\frac{C_1}{\hbar}\frac{1}{\om_1+\om}+\frac{2C_2V_{12}}{\hbar\om}\frac{\om_2}{\om_1^2-\om_2^2},
\label{hh7}
\ee
since $\sig_{10}(t)|\psi\ra=0$,  $\la a_0(t)a_0(t)\ra=\la\ad_0(t)a_0(t)\ra=0$, and $\la a_0(t)\ad_0(t)\ra=1$, and
\be 
\la H_c^{(1)}\ra\cong -\frac{2C_1^2/\hbar}{\om_1+\om}-\frac{2C_1C_2V_{12}}{\hbar\om}\frac{\om_2}{\om_1^2-\om_2^2}.
\label{hh8}
\ee
The same approach for atom 2 gives
\be
\la H_c^{(2)}\ra\cong -\frac{2C_2^2/\hbar}{\om_2+\om}-\frac{2C_1C_2V_{12}}{\hbar\om}\frac{\om_1}{\om_2^2-\om_1^2}.
\label{hh9}
\ee

For the expectation value of $H_c=H_c^{(1)}+H_c^{(2)}$ we therefore obtain
\be
\la H_c\ra\cong-\frac{2C_1^2/\hbar}{\om_1+\om}-\frac{2C_2^2/\hbar}{\om_2+\om}-\frac{2C_1C_2V_{12}/\hbar\om}{\om_1+\om_2}.
\label{h10}
\ee
Since $C_1^2$ and $C_2^2$ are proportional to $e^2$, and $C_1C_2V_{12}$ is proportional to $e^4$, the PHF theorem introduces factors
\be
\int_0^1\frac{d\lambda}{\lambda}\lambda^2=\frac{1}{2} \ \ \ \ {\rm and} \ \ \ \ \int_0^1\frac{d\lambda}{\lambda}\lambda^4=\frac{1}{4}
\label{h11}
\ee 
for the first two terms and the last term, respectively, on the right-hand side of the expression (\ref{h10}), so that the change in the atom-field system due to their interaction is
\be
\Delta E=-A_1^2(\bd_1\cdot\hat{\bf e})^2\frac{\om}{\om+\om_1}-A_2^2(\bd_2\cdot\hat{\bf e})^2\frac{\om}{\om+\om_2}
-\frac{1}{2}\frac{A_1A_2}{r^3}\frac{(\bd_1\cdot\hat{\bf e})(\bd_2\cdot\hat{\bf e})[\bd_1\cdot\bd_2-3(\bd_1\cdot\hat{\bf e})(\bd_2\cdot\hat{\bf e})]}{\om_1+\om_2}
\label{h12}
\ee
when we use the definition (\ref{haug4}).

The two-state model simplifies some algebra in our Heisenberg-picture calculation, but the result (\ref{h12}) is easily generalized to include contributions from all the allowed transitions from the ground states of the two atoms. The first term on the right-hand side of Eq. (\ref{h12}), for instance, generalizes to
\be
\Delta E_1=-A_1^2\sum_{s}|(\bd_1\cdot\hat{\bf e})_{sg}|^2\frac{\om}{\om+\om_{sg}}
\label{h13}
\ee
in the notation of Eq. (\ref{eq18}). After accounting for additional, self-energy terms, we obtain the Lamb shift due to coupling of atom 1 to the single-mode field. But of greater interest here is the interatomic interaction term in the expression (\ref{h12}): for strong coupling to a single-mode field the nonretarded van der Waals interaction varies as $1/r^3$ rather than $1/r^6$ \cite{haugland,philbin,remarkk}. When generalized to include all allowed transitions from the ground states, we obtain, except for the factor $A_1A_2$, the result of the perturbation-theoretic analysis of Reference \cite{philbin}. This factor has an interesting implication for the physical interpretation of the $1/r^3$ interaction, as discussed in the following section. 

\section{\label{sec:remarks}Remarks}
Zero-temperature Lamb shifts and van der Waals interactions have clear physical interpretations in terms of fluctuating zero-point fields. In particular, for the van der Waals interaction between two atoms in free space, each atom is ``driven" by the zero-point field at its location, and the fluctuations of the zero-point fields at the two locations are correlated. The correlation falls off rapidly with the distance $r$ between the two locations, giving the $r^{-6}$ dependence of the nonretarded van der Waals interaction. 

In the case of strong coupling of the atoms to a single cavity mode, unlike the case in which the atoms are coupled to the infinite set of modes of free space, there is no decrease in electric field correlations with $r$, and for small $r$ the van der Waals interaction varies as $r^{-3}$ rather than $r^{-6}$. Such $r^{-3}$ behavior is also found  in a different scenario, when each atom experiences an {\sl externally applied} single-mode field \cite{pwmsmith}. In this case the interpretation of the $r^{-3}$ behavior is obvious: the two atoms have correlated induced dipole moments and experience, for small $r$, just the $r^{-3}$ electrostatic dipole-dipole interaction. But in the more subtle $r^{-3}$ behavior resulting from coupling of the atoms to a {\sl zero-point}, vacuum cavity mode \cite{haugland,philbin}, each atom has a dipole moment induced by the zero-point field whose fluctuations are correlated for effectively all values of $r$. We note that the presence of the factor $A_1A_2$ in the energy (\ref{h12}) implies that there is no van der Waals interaction if one of the atoms finds itself at a node of the cavity field, i.e., if either $A_1$ or $A_2$ vanishes.

The derivations of the zero-temperature Lamb shifts and van der Waals interactions using the PHF theorem make it clear that these effects are attributable to the fluctuations of the zero-point electromagnetic field. They can also be said to be attributable to changes in zero-point energy, as in Feynman's argument for the Lamb shift in Section \ref{sec:energy}. But the fluctuation perspective seems to offer a more physical picture of interacting dipoles as opposed to just energy ``bookkeeping." Moreover, Lamb shifts and van der Waals interactions can be understood from the perspective of the quantum fluctuations not of zero-point fields but of the ``source" fields generated by the dipoles themselves. The same is true of Casimir's famous attraction between conducting plates \cite{jaffe}.

\section*{ACKNOWLEDGMENTS}
It is a pleasure to submit this paper in recognition of the creative and influential contributions of Professor Iwo Bialynicki--Birula. I thank P. R. Berman and G. W. Ford for helping me to better understand the ``remarkable formula" and van der Waals interactions, and G. J. Maclay for informative discussions relating to Lamb shifts.

\appendix*
\section{Electric field correlations and dyadic Green function}
The positive-frequency part of the electric field operator for a vacuum or thermal field can be expressed as
\be
\bE^{(+)}(\br,t)=i\sum\rmk\Big(\frac{2\pi\hbar\om_k}{V}\Big)^{1/2}a\rmk e^{-i\om_kt}e^{i\bk\cdot\br}\hat{\bf e}\rmk,
\label{eqa}
\ee
where as usual $a\rmk$ is the photon annihilation operator for the plane-wave mode with wave vector $\bk$ and polarization index $\lambda$. For thermal radiation, $\la\ad\rmk a_{\bk\pr\lambda\pr}\ra=q(\om)\delta^3_{\bk\bk\pr}\delta_{\lambda\lambda\pr}$ and 
$\la a\rmk \ad_{\bk\pr\lambda\pr}\ra=[q(\om)+1]\delta^3_{\bk\bk\pr}\delta_{\lambda\lambda\pr}$, $q(\om)=(e^{\hbar\om/k_BT}-1)^{-1}$, and if follows after taking $\sum\rmk(...)\rightarrow V/(2\pi)^3\sum_{\lambda}\int d^3k(...)$ in the familar fashion that
\bea
\la \Ep_i(\br_n,t)\Em_j(\br_m,t\pr)\ra&=&\frac{\hbar}{\pi c^3}\int_0^{\infty}d\om\om^3[q(\om)+1]F_{ij}(\om r/c)e^{i\om(t\pr-t)},\nonumber\\
\la \Em_i(\br_n,t)\Ep_j(\br_m,t\pr)\ra&=&\frac{\hbar}{\pi c^3}\int_0^{\infty}d\om\om^3q(\om)F_{ij}(\om r/c)e^{-i\om(t\pr-t)}, \ \ \ r=|\br_n-\br_m|,
\label{eq22}
\eea
\be
F_{ij}(x)\equiv(\delta_{ij}-\hat{\br}_i\hat{\br}_j)\frac{\sin{x}}{x}+(\delta_{ij}-3\hat{\br}_i\hat{\br}_j)\big(\frac{\cos{x}}{x^2}-\frac{\sin{x}}{x^3}\big).
\label{eq23}
\ee
Thus, for thermal radiation,
\bea
\big\la \Ep_i(\br_n,\om)\Em_j(\br_m,\om\pr)\big\ra&=&\frac{\hbar}{\pi c^3}\om^3[q(\om)+1]F_{ij}(\om r/c)\delta(\om-\om\pr),\nonumber\\
\big\la \Em_i(\br_n,\om)\Ep_j(\br_m,\om\pr)\big\ra&=&\frac{\hbar}{\pi c^3}\om^3q(\om)F_{ij}(\om r/c)\delta(\om-\om\pr).
\label{eq24}
\eea

The electric field $\bE_S(\br_n,t)$ at a point $\br_n$ from an electric dipole source at $\br_m$ is
\be
E_{Si}(\br_n,t)=-\frac{1}{c^2r}(\delta_{ij}-\hat{\br}_i\hat{\br}_j)\ddot{p}_j(t-r/c)-(\delta_{ij}-3\hat{\br}_i\hat{\br}_j)\big[\frac{1}{cr^2}\dot{p}_j(t-r/c)+
\frac{1}{r^3}p_j(t-r/c)\big].
\label{eq25}
\ee
We therefore identify
\be
\Ep_{Si}(\br_n,\om)=G_{ij}(\br_n,\br_m,\om)\ppp_j(\br_m,\om),
\label{eq26}
\ee
\be
G_{ij}(\br_n,\br_m,\om)=k\frac{\om^2}{c^2}\Big[(\delta_{ij}-\hat{\br}_i\hat{\br}_j)\frac{1}{kr}+(\delta_{ij}-3\hat{\br}_i\hat{\br}_j)\big(\frac{i}{k^2r^2}-\frac{1}{k^3r^3}\big)
\Big]e^{ikr}, \ \ \ k=n\om/c,
\label{eq27}
\ee
and Eqs. (\ref{eq19}) then follow from Eqs. (\ref{eq24}).


\begin{thebibliography}{99}
\bibitem{remark2} We do not include any line broadening effects that would give an imaginary part to the polarizability. In particular, in our formulation radiation reaction is accounted for in Eq. (\ref{ad1}) but not in Eq. (\ref{eq18}). But the polarizability must not have any poles in the upper half of the complex frequency plane, whence we add $i0^+$ to $\om$ in the Kramers--Heisenberg formula.
\bibitem{remarkk} When we include a term of second order in $V_{12}$ in our calculation we obtain another contribution to $\Delta E$ that corresponds to the familiar $1/r^6$ van der Waals interaction.
\end{thebibliography}
\end{document}